\documentclass[aps,prb,twocolumn,showpacs,superscriptaddress,floatfix]{revtex4-1}

\pdfoutput=1
\usepackage[titletoc,toc,title]{appendix}
\usepackage{amsmath,amssymb}
\usepackage{graphicx}
\usepackage{color}
\usepackage{rotating}
\usepackage{bm}
\usepackage{setspace}
\usepackage{hyperref}
\usepackage{float}
\usepackage{soul}
\usepackage[T1]{fontenc}
\usepackage{bbold}
\usepackage{color}
\usepackage{braket}

\hypersetup{
colorlinks=true,
urlcolor= blue,
citecolor=blue,
linkcolor= blue,
bookmarks=true,
bookmarksopen=false,
}

\renewcommand{\vec}[1]{\bm{#1}}

\begin{document}

\preprint{APS/123-QED}

\title{Duality between atomic configurations and Bloch states in twistronic materials.}
\author{Stephen Carr}
\affiliation{Department of Physics, Harvard University, Cambridge, Massachusetts 02138, USA}
\author{Daniel Massatt}
\affiliation{Department of Statistics, University of Chicago, Chicago, Illinois 60615, USA}
\author{Mitchell Luskin}
\affiliation{School of Mathematics, University of Minnesota, Minneapolis, Minnesota 55455, USA}
\author{Efthimios Kaxiras}
\affiliation{Department of Physics, Harvard University, Cambridge, Massachusetts 02138, USA}
\affiliation{John A. Paulson School of Engineering and Applied Sciences, Harvard University, Cambridge, Massachusetts 02138, USA}

\date{\today}

\begin{abstract}
	
The relative orientation (twist) of successive layers of stacked two-dimensional (2D) materials creates variations in the interlayer atomic registry.
The variations often form a super lattice, called a moir\'e pattern, which can alter electronic properties.
In this work we introduce a classification of the single-particle electronic structures that can occur in twisted stacks of 2D layers by characterizing them as ``moir\'e molecules'' or ``moir\'e crystals''.
The molecules generate localized electronic states and moir\'e flat bands, while the crystals are sometimes unconventional and produce electronic banding in the configuration basis.
The underpinning of this classification is the duality between interlayer configuration and monolayer Bloch momentum in moir\'e Hamiltonians.
We apply this understanding to diagrams of local electron density in untwisted geometries
to produce intuitive and quantitative predictions of twistronic properties.
We provide a conceptual introduction to 
this framework through a one-dimensional model, 
and then apply it to 2D twisted bilayers of the semi-metal graphene, 
and of MoS$_2$, a representative material of the transition metal 
dichalcogenide (TMDC) family of semiconductors.
This level of thorough understanding of twistronic phenomena 
is vital in the search for new material platforms for localized moir\'e electrons.
\end{abstract}

\maketitle

\section{Introduction}
\label{sec:intro}

    The number of two-dimensional (2D) materials that 
    have been experimentally isolated as single-layers of atomic-scale thickness has been  
    growing at a rapid pace \cite{Novoselov2004, Ayari2007, Dean2010, Mak2010, Radisavljevic2011, De2013}. 
    Alongside these advances, methods to combine single layers
    into multi-layer heterostructures \cite{Hunt2013,Cao2016,Koren2016,Kim2017} have developed a 
    remarkable level of control over the relative orientation (twist) of successive layers, 
    on the order of $0.1^\circ$.
    The resulting heterostrucures represent a new type of composite materials 
    with properties spanning a vast range that includes insulators, semiconductors, 
    metals and superconductors \cite{Geim2013}.
    The different arrangements of layers of 2D materials
    provide an intriguing platform for exploring new physics and potential applications
    based on their electronic, optical, magnetic and thermal properties.
    The relative orientation of successive layers, often characterized by a twist 
    angle between the ideal in-plane lattices, represents an additional ``knob'' for 
    adjusting the system properties. 
    A well-studied example is twisted bilayer graphene (tBLG), where the twist-angle between successive layers of graphene causes controllable inter-layer electronic hybridization.
    At a magic twist angle ($\approx 1^\circ$) a symmetric hybridization between the two Dirac cones from the individual layers introduces flat
    bands in the bilayer band structure, resulting in localized electronic states \cite{Li2009, Bistritzer2011, San-Jose2012, Wong2015, Brihuega2012, Luican2011} and enhanced electronic correlations \cite{Cao2018sc, Cao2018mott, Yankowitz2019}.
    
    In the study of twisted bilayers, the use of a supercell approximation \cite{Uchida2014, Matsushita2018, Naik2018, Alessandro2019, Xian2019} or a continuum model \cite{Santos2007,Bistritzer2011,Lopes2012,Wu2018} imposes a periodic length-scale for the system to provide interpretable band structures.
    We show that instead of relying on the bands of Bloch theory in momentum space, the local density of states (LDoS) in either the space of atomic configurations or Bloch states provides a surprising amount of clarity in the study of twisted electronic structure.
    The experimentally important notions of flat bands and moir\'e band gaps are still obtainable in this context.
    In addition, the LDoS can form sharply defined features in configuration space and energy, which we call ``configuration banding''.
    These two regimes of interesting twistronic features, hosting localized modes or banding, are analogous to the electronic structure observed in conventional molecules (localized states) or crystals (extended states or bands).
    This pattern arises because of a duality between the momentum and position operators that can occur in specific scenarios for moir\'e Hamiltonians.
 
    To illustrate the utility of the duality interpretation in a simple context, we first examine a one-dimensional incommensurate chain model.
    Moir\'e flat bands occur when only a finite number of Bloch states or local configurations are needed to capture how two parabolic band extrema are coupled through the interlayer coupling; 
    this corresponds to the ``moir\'e molecule''.
    Configuration banding occurs when an infinite number of Bloch states are needed to accurately capture the interlayer interaction between two band structures; this corresponds to 
    the ``momentum crystal'', which often exists alongside conventional Bloch bands in the moir\'e systems.
    Expanding on the simple 1D example, we then investigate 
    similar features in the electronic structure of realistic materials, such as graphene and MoS$_2$, a representative of the transition metal dichalcogenide family.
    The moir\'e molecule interpretation can be modeled by a harmonic oscillator and allows for accurate prediction of twist-induced flat bands.
    The momentum crystal interpretation can also provide highly localized electronic states, and provides an alternate explanation of the localization caused by strong incommensurate potentials.
    
    This paper is organized as follows: Section \ref{sec:methods} contains a discussion of the methods used for modeling electronic structure of 1D and 2D moir\'e systems, Section \ref{sec:1d} categorizes twistronic spectral features for 1D systems, Section \ref{sec:2d} applies this categorization to realistic 2D moir\'e bilayers, and Section \ref{sec:conc} summarizes the results.
    The appendix examines how configuration banding manifests in the real space LDoS.

	\begin{figure*}
		\centering
		\includegraphics[width=1\textwidth]{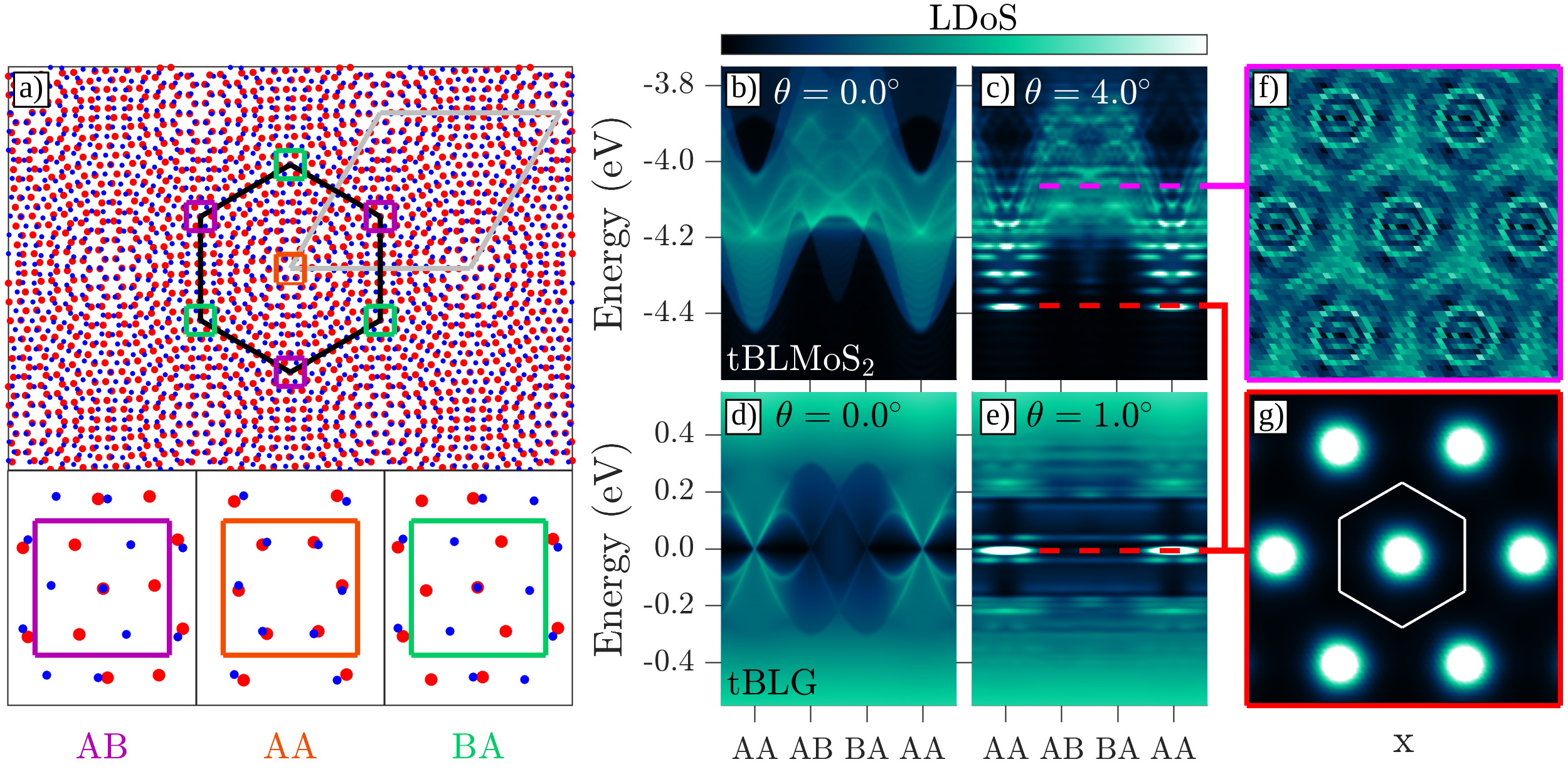}
		\caption{
    		\textbf{a)} Geometry of a twisted bilayer formed by two identical honeycomb lattices. The bottom (top) layer is red (blue).
    		The moir\'e supercell is shown in grey and the moir\'e Wigner-Seitz cell is outlined in black, with colored boxes highlighting the high-symmetry stacking locations shown in the insets, corresponding to $AB$, $AA$, and $BA$ stackings.
    		\textbf{b)} Configuration-dependent local density of states (LDoS) for $p_z$ orbitals in untwisted bilayer MoS$_2$ in an energy region 4 eV  below the valence band maximum.
    		\textbf{c)} Configuration-dependent LDoS for a $4^\circ$ twisted bilayer of MoS$_2$.
    		\textbf{d-e)} The same as \textbf{(b-c)}, but for twisted bilayer graphene at $0.0^\circ$ and $1.0^\circ$.
    		\textbf{f-g)} Top-down views of the real space variation in the LDoS at the energy values identified by the dashed colored lines.
    		The LDoS in \textbf{(g)} appears in both tBLG and tBLMoS$_2$, and includes the moir\'e Wigner-Seitz cell outlined in white.}
		\label{fig:ldos_intro}
	\end{figure*}

\section{Methods}
\label{sec:methods}
    
    It is a daunting computational task to calculate accurately the properties of layered structures that involve a small twist angle.
    This is because these atomic arrangements are either incommensurate or have periodicity 
    over length scales several orders of magnitude larger than the primitive unit cells of the individual layers.  
    In tBLG, the Dirac cone yields a simple low-energy Hamiltonian from which momentum-scattering interactions between layers can be computed \cite{Santos2007, Mele2010, Bistritzer2011, Mele2012, Lopes2012, Moon2013}.
    For transition-metal dichalcogenides (TMDC) similar models for the parabolic band extrema have been used \cite{Kormanyos2015, Wu2017, Wu2018}.
    Here, we employ accurate \textit{ab-initio} tight-binding hamiltonians \cite{Fang2015, Fang2016} derived from 
    density functional theory (DFT) and maximally localized Wannier functions (MLWF) \cite{Marzari2012} within a framework specifically developed for twisted 2D systems \cite{Massatt2017, Carr2017}.
    
    The tight-binding modeling makes it feasible to investigate the electronic features in twisted 2D materials in two different ways.
    First, by calculating the band structure through diagonalization of twisted supercell Hamiltonians.
    Second, by calculating the local density of states (LDoS) of aperiodic systems by computing local spectral properties at the center of large finite 
    regions with the kernel polynomial method (KPM) \cite{Weisse2006a, Carr2017}.
    For the latter method, we used circular disks of diameter of $80$ nm for TMDCs and of $200$ nm for graphene, with both geometries containing millions of atomic orbitals.
    
    In Fig. \ref{fig:ldos_intro} we introduce an important concept for exploring the physics, namely displaying the LDoS for atomic structures chosen along lines in configuration space which connect the high symmetry stackings of the bilayer.
    This is analogous to displaying band-structures in reciprocal (momentum) space 
    along lines connecting high-symmetry points in the Brillouin Zone (BZ).
    The Wigner-Sietz cell of the moir\'e supercell is outlined in Fig. \ref{fig:ldos_intro}a, with insets showing the local atomic configuration at selected points.
    The LDoS results displayed in Fig. \ref{fig:ldos_intro}b-g) are not generated by looking at different atoms in the same structure.
    Each configuration is calculated independently of all the others by uniformly shifting the top layer relative to the bottom one.
    
    For both tBLG and twisted bilayer MoS$_2$ (tBLMoS$_2$) near $0^\circ$, there are three high symmetry stacking configurations: $AA$, $AB$, and $BA$.
    The sublattices of the honeycomb structure, $A$ and $B$, are used to describe the alignment between the two layers.
    $AA$ has both the $A$ and $B$ atoms aligned vertically, while $AB$ and $BA$ have only one pair of opposite sublattice atoms aligned vertically between layers.
    For graphene, both sublattices are identical carbon atoms, while the hexagonal TMDC hosts a metal on the $A$ site and a dimer of chaclogen atoms on the $B$ site.
    Note that the $AA$ and $AB$ notation is often used in the TMDCs to instead distinguish between the two distinct phases of an aligned bilayer: the $0^\circ \pm 120^\circ$ alignment and $60^\circ \pm 120^\circ$.
    Although near $0^\circ$ twist the TMDC bilayers have the same symmetry as tBLG (with $AB$ and $BA$ stacking related), due to the non-identical sublattices that symmetry is broken for TMDC bilayers near $60^\circ$ twist.

    The smooth variations in LDoS at $\theta = 0^\circ$ can be attributed solely to the local variation in interlayer coupling over configuration space \cite{Zhang2017}.
    For $\theta \neq 0^\circ$, localized modes appear in both tBLG and tBLMoS$_2$ (Fig. \ref{fig:ldos_intro}g). 
    Their appearance in tBLG at $\theta = 1.0^\circ$ corresponds to the localized wavefunction caused by moir\'e flat bands at the so-called magic angle \cite{Bistritzer2011}.
    The tBLMoS$_2$ LDoS of Fig. \ref{fig:ldos_intro}f is more complicated.
    The curved and straight features in the LDoS smoothly expand or contract as the energy changes, caused by surfaces of high electron density in the configuration-energy space.
    This is an example of electronic banding in the configuration basis.

\section{One-dimensional model}
\label{sec:1d}    
    
\begin{figure}
    		\centering
    	    \includegraphics[width=0.5\textwidth]{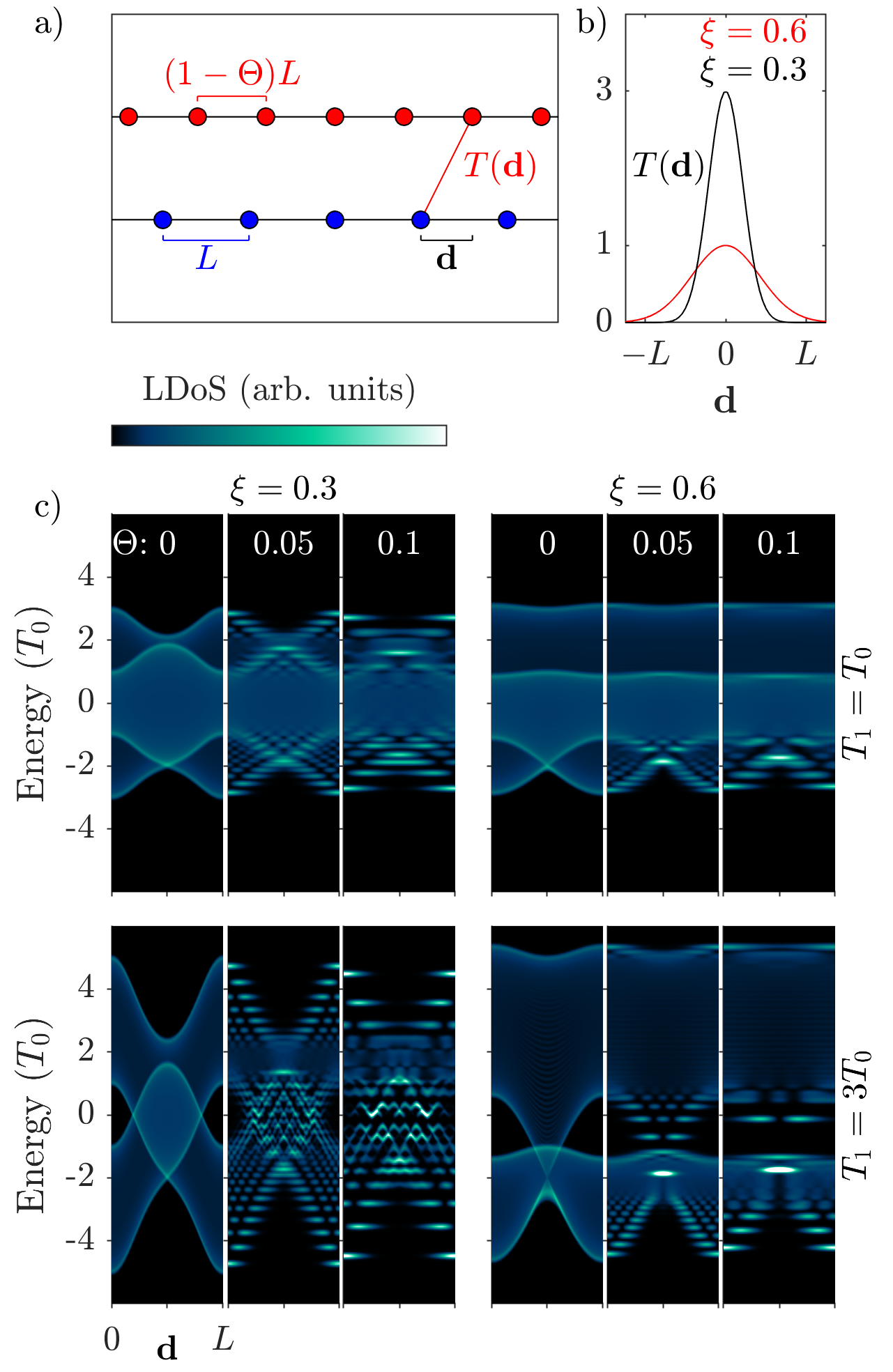}
    		\caption{
    		\textbf{a)} The 1D two-chain model, with the top (bottom) chain in red (blue) with its lattice parameter in the same color.
    		The relative interlayer configuration between the top and bottom layer ($\mathbf{d}$) and its interlayer coupling ($T(\mathbf{d})$) are labeled for a pair of atoms. 
    		\textbf{b)} The interlayer coupling function $T$, with parameters $\xi = 0.3$ $(0.6)$ and $T_1 = 3$ $(1)$ in black (red).
    		\textbf{c)} Configuration-dependent LDoS of the top atom in the two-chain model.
    		Each of the four panels consists of an ``untwisted'' geometry with $\Theta = 0$, and  two ``twisted'' geometries with $\Theta = 0.05$ and $0.1$, for two different values of $\xi$, 
    		$\xi = 0.3$ (left set) and $\xi=0.6$ (right set), and two different values of $T_1$, $T_1=1$ (top set) 
		and $T_1 = 3$ (bottom set).
    		}
    		\label{fig:1d_overview}
    	\end{figure}

    To investigate the nature of twistronic phenomena, we consider a simple 1D model 
    consisting of two chains of single-atom unit cells with a starting lattice parameter of $L = 1$.
    The mismatch is introduced by changing the lattice spacing of one layer to $(1 - \Theta) L$, for a small $\Theta$ value as shown in Fig. \ref{fig:1d_overview}(a). 
    $\Theta$ encodes a lattice mismatch between the chains, and is the 1D equivalent to the twist angle.
    Although we focus on twisted structures in the case of 2D bilayers, moir\'e superlattices generated by a small lattice mismatch show similar phenomena.
    
    For the single layer electronic structure, a nearest-neighbor hopping of strength $T_0 = 1$ defines the characteristic energy scale for the system.
    With two atoms in different layers separated by an in-plane distance $\mathbf{d}$, the interlayer coupling is 
    	\begin{equation}
    	T(\mathbf{d}) = T_1  e^{-(d/\xi)^2}.
    	\end{equation}
    The local density of states (LDoS) for this model for various choices of $\Theta$, $\xi$, and $T_1$ is shown in Fig. \ref{fig:1d_overview}.
    This LDoS is calculated by the conventional method: diagonalizing a periodic supercell with $10$ $k$-point samples in the 1D Brillouin zone.

    Focusing on the $\Theta = 0$ cases, there are a number of common features.
    Near the extrema of the electronic states, bright lines correspond to parabolic extrema in the one-dimensional band structure.
    The energy of the band extremum depends on the interlayer stacking, $\mathbf{d}$, because of the stacking-dependent interlayer coupling.
    In this simple system, this is caused by a splitting of the two monolayers' identical spectra by the interlayer coupling strength, creating copies of the monolayer band structure that are either shifted down or up in energy (corresponding to bonding/anti-bonding combinations).

    When the atoms of both layers are on top of each other ($\mathbf{d} = 0$ or $L$) the interlayer coupling is strongest, causing the largest splitting of the two copies.
    In the two $\xi = 0.3$ cases, the splitting is minimal as the layers have almost no interlayer coupling at $\mathbf{d} = L/2$.
    Although a larger value of $\xi$ (longer range coupling) will generally increase the absolute strength of the interlayer coupling for a given stacking, it reduces the variation in the interlayer coupling as it begins to average over adjacent sites.
    This is visible when comparing the $\mathbf{d}$ variation in the positive LDoS extrema between $\xi = 0.3$ and $0.6$ calculations.
    Increasing the variation in the interlayer coupling, and not its absolute strength, is the key to inducing moir\'e phenomena, so generally a small $\xi$ is preferred.
    
    The LDoS of the $\Theta \neq 0$ systems fall into three general categories.
    Near band extrema isolated bright spots can occur, corresponding to the emergence of moir\'e flat bands.
    Away from band extrema two scenarios are possible.
    The LDoS can either be featureless in both energy and $\mathbf{d}$, or it can have sharply defined lines criss-crossing one another.
    The former is the signature of conventional Bloch waves, while the later is a unique feature of the moir\'e systems, the phenomenon we defined as \textit{configuration banding}.
    In the following subsections we will investigate each of these categories with the 1D model, explaining how they arise and illustrating the duality between the configuration and Bloch bases in twistronic materials.
    
\subsection{Flat bands from a harmonic approximation}
    
\begin{figure}
      \centering
      \includegraphics[width=0.5\textwidth]{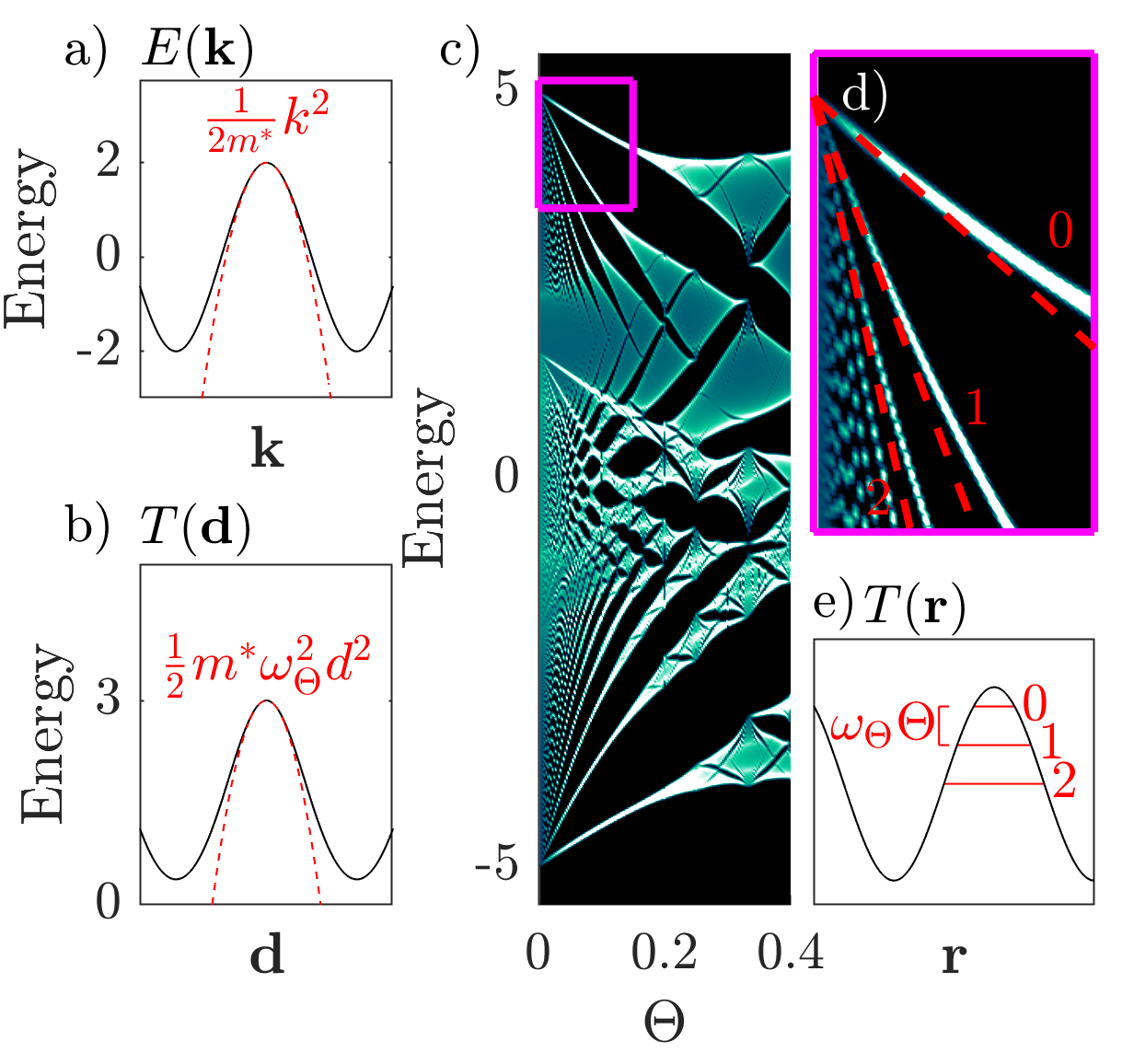}
      \caption{
            \textbf{a)} Band structure of a single 1D chain, with a harmonic approximation of a band extremum in red.
            \textbf{b)} Configuration dependence of the interlayer coupling function for the bilayer 1D chain model, with a harmonic approximation of its maximum in red.
            \textbf{c)} Total density of states (DoS) for commensurate bilayer chains for $\Theta$ in $[0, 0.4]$.
            A fractal butterfly spectrum is evident, with a region highlighted in the top left corner.
            \textbf{d)} Expanded view of the region highlighted in (c), 
            that shows a number of isolated $\Theta$-dependent states in the DoS.
            The dashed red lines correspond to the energy levels, $\omega (n + 1/2)$ of a real space harmonic oscillator based on the red fits in \textbf{(a,b)}.
            The red numbers indicate the $n$ for that level.
            \textbf{e)} Schematic of the fitted real space harmonic well, with the first three energy levels of the oscillator indicated by the red lines and numbers.
            The frequency $\omega = \omega_\Theta \Theta$ is displayed between the first two energy levels.
           }
      \label{fig:1d_qho}
    \end{figure}

    The moir\'e flat-bands can be explained in this model by performing a perturbative expansion in both the monolayer eigenvalues, $E(\mathbf{k})$, and the interlayer coupling, $T(\mathbf{d})$.
    As we are concerned only with the band extremum in this case, the monolayer Hamiltonians can be modeled as a uniform electron gas with effective mass $m^*$  
    \begin{equation}
    H_i = \frac{\sigma}{2m^*} (\mathbf{k}-\mathbf{k}^i_0)^2
    \end{equation}  
    where $\sigma = \pm 1$ depends on if we are interested in a valence (hole) band maximum or conduction (electron) band minimum.
    The interlayer coupling can be approximated as a perturbing potential in space, $T(\vec{r})$.
    When the band extremum is located at the $\Gamma$ point ($\mathbf{k}^i_0 = 0$), 
    the full Hamiltonian can be written entirely in the position basis:   
    \begin{equation}
    H = \frac{\sigma}{2m^*} (\psi_1^\dagger \partial^2 \psi_1 + \psi_2^\dagger \partial^2 \psi_2) + T(\vec{r}) (\psi_1^\dagger \psi_2 + h.c.).
    \end{equation}
    Defining both a bonding and an anti-bonding pairing between the layers, $\psi_\pm = \frac{1}{\sqrt{2}} \left( \psi_1 \pm \psi_2 \right)$, separates this into two single-variable Hamiltonians:
    \begin{equation}
    \label{eq:psi_pm_model}
    H_\pm = \frac{\sigma}{2m^*} (\psi_\pm^\dagger \partial^2 \psi_\pm) \pm T(\vec{r}) \psi_\pm^\dagger \psi_\pm.
    \end{equation} 
    If the interlayer coupling can be locally expanded at its extrema in a quadratic form $T(\vec{r}) = \frac{\sigma}{2} m^* \omega^2 r^2$, one obtains the harmonic oscillator equation
    \begin{equation}
    H_\pm =  \sigma \psi_\pm^\dagger \left( \frac{1}{2m^*} \partial^2 \pm \frac{1}{2} m^* \omega^2 r^2 \right) \psi_\pm.
    \end{equation}
    This concept is illustrated in Fig. \ref{fig:1d_qho}a,b, for a hole-like band ($\sigma = -1$), and the effective parameters $m^*$ and $\omega$ are easily extracted from the simple 1D model.
    The band structure for a 1D chain with nearest neighbor coupling $T_0$ and lattice parameter $L$ is given by $2 T_0 \cos{k L} \approx 2 T_0 + T_0 L^2 k^2$.
    Assuming $\Theta$ is small, we can take $m^* = (2 T_0 L^2)^{-1}$ for both chains.
    Here we have kept terms of $T_0$ and $L$ (both equal to $1$) to ensure a general result.
    The value of $\omega$ depends on both the strength of interlayer coupling, $T_1$, and the moir\'e length, $\lambda = 1/\Theta$.
    Making the substitution $\mathbf{d} = \mathbf{r}/\lambda$ in the definition of $T(\vec{d})$ gives
    
    \begin{equation}
T(\vec{r}) \approx T_1 e^{-\left( \frac{r}{\lambda \xi}\right)^2} \approx T_1 \left( 1 - \frac{\Theta^2 r^2}{\xi^2} \right).
    \end{equation}
    
    Thus $\frac{1}{2} m^* \omega^2 = \Theta^2 T_1 / \xi^2$, or $\omega = 2 L \Theta \sqrt{T_0 T_1} / \xi = \omega_{\Theta} \Theta$.
    The $\Theta$-independent part of the above expression, $\omega_{\Theta}$, is a useful parameter when analyzing moir\'e flat bands that arise from parabolic extrema.
    
    Comparing the expected energy levels of the harmonic oscillator, $(\frac{1}{2} + n) \omega$, to the calculated DOS in Fig. \ref{fig:1d_qho}d  shows good agreement for small $\Theta$ ($\le 0.1$).
    For larger $\Theta$ values, the electronic density fans away from the ideal oscillator modes as the width of the associated moir\'e bands grows.
    For ideal ``flatness'' of the $n^\textrm{th}$ oscillator state, we find the condition to be $(n + 2.5) \omega < T_1$, or at least three harmonic oscillator states must fit in the interlayer potential well, as shown in Fig. \ref{fig:1d_qho}e.
    
\subsection{Moir\'e Molecules}
    
\begin{figure}
      \centering
      \includegraphics[width=0.5\textwidth]{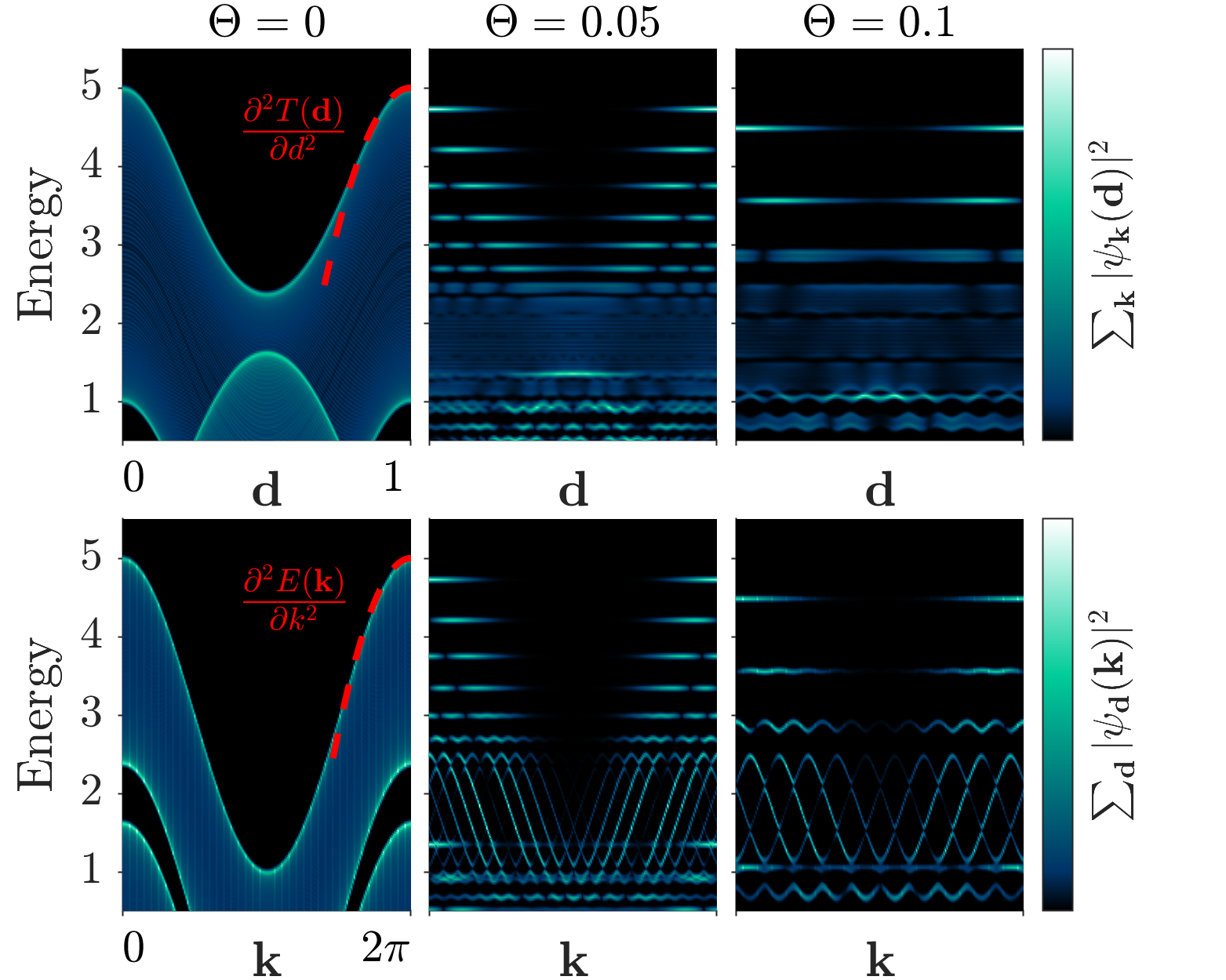}
      \caption{
            Electronic structure calculations for bilayer chains with $T_1 = 3$ and $\xi = 0.3$, with varying $\Theta$.
            \textbf{Top:} LDoS for atoms of the top layer with specific configurations $\mathbf{d}$.
            The harmonic approximation of $T(\mathbf{d})$ is highlighted in red.
            \textbf{Bottom:} LDoS for Bloch waves of the top layer with specific momenta $\mathbf{k}$.
            The harmonic approximation of the bands $E(\mathbf{k})$ is highlighted in red.
           }
      \label{fig:duality_molecule}
    \end{figure}

    Beyond its use to predict moir\'e flat bands, the harmonic oscillator equation also serves as a clear example of momentum-configuration duality.
    The two terms in the harmonic equation, quadratic in either the momentum operator or the position operator, are interchangeable provided their prefactors are comparable.
    For the momentum term, the prefactor was simply the curvature of the electronic bands (i.e. $1/m^*$):
    \begin{equation}
    \frac{\partial^2 E(\mathbf{k})}{\partial \mathbf{k}^2}.
    \end{equation}
    For the position term, it was the curvature of the interlayer coupling potential 
    \begin{equation}
    \frac{\partial^2 T(\mathbf{r})}{\partial \mathbf{r}^2} = \Theta^2  \frac{\partial^2 T(\mathbf{d})}{\partial \mathbf{d}^2}.
    \end{equation}
    Putting these together, the frequency of the harmonic oscillator is
    \begin{equation}
    \omega = \omega_\Theta \Theta =
    \Theta 
    \sqrt{ \frac{\partial^2 E(\mathbf{k})}{\partial \mathbf{k}^2}
    \frac{\partial^2 T(\mathbf{d})}{\partial \mathbf{d}^2} }.
    \end{equation}
    This illustrates the dual nature of momentum and configuration spaces in moir\'e problems.
    Indeed, we can re-examine the LDoS calculation to see the duality more clearly.
    In configuration space, we computed the LDoS as a function of configuration, but implicitly summed over all the momentum degrees of freedom for the moir\'e supercell:
    \begin{equation}
    \textrm{LDoS}(\mathbf{d}) = \sum_\mathbf{k} |\psi_{\mathbf{k}}(\mathbf{d})|^2
    \end{equation}
     Reformulating this equation under the duality, we can instead calculate the LDoS in momentum space by summing over all configurations for a specific moir\'e wavenumber:   
    \begin{equation}
    \textrm{LDoS}(\mathbf{k}) = \sum_\mathbf{d} |\psi_{\mathbf{d}}(\mathbf{k})|^2
    \end{equation}    
    In Fig. \ref{fig:duality_molecule}, this is presented for the positive band extremum of the 1D model.
    Clearly defined states are isolated in energy in both the configuration and momentum bases.
    If the curvature of $E(\mathbf{k})$ is equal to that of $T(\mathbf{d})$, the two sets of figures would be indistinguishable near the band extremum.
    In both configuration or momentum space, a strong dependence of the spectrum on the local variable ($\mathbf{d}$ or $\mathbf{k}$) causes pockets of isolated eigenvalues in the energy spectrum.
    These couple to one another, forming an effective molecule and creating electronic structure reminiscent of molecular orbitals in both configuration and momentum space.
    The higher-energy harmonic states have more nodal points, corresponding to higher ``momentum'' in 1D.
    This concept generalizes to 2D crystals, where the higher energy moir\'e molecular states have larger angular momentum ($s$-wave, $p$-wave, etc.)

\subsection{Moir\'e Crystals}

\begin{figure}
      \centering
      \includegraphics[width=0.5\textwidth]{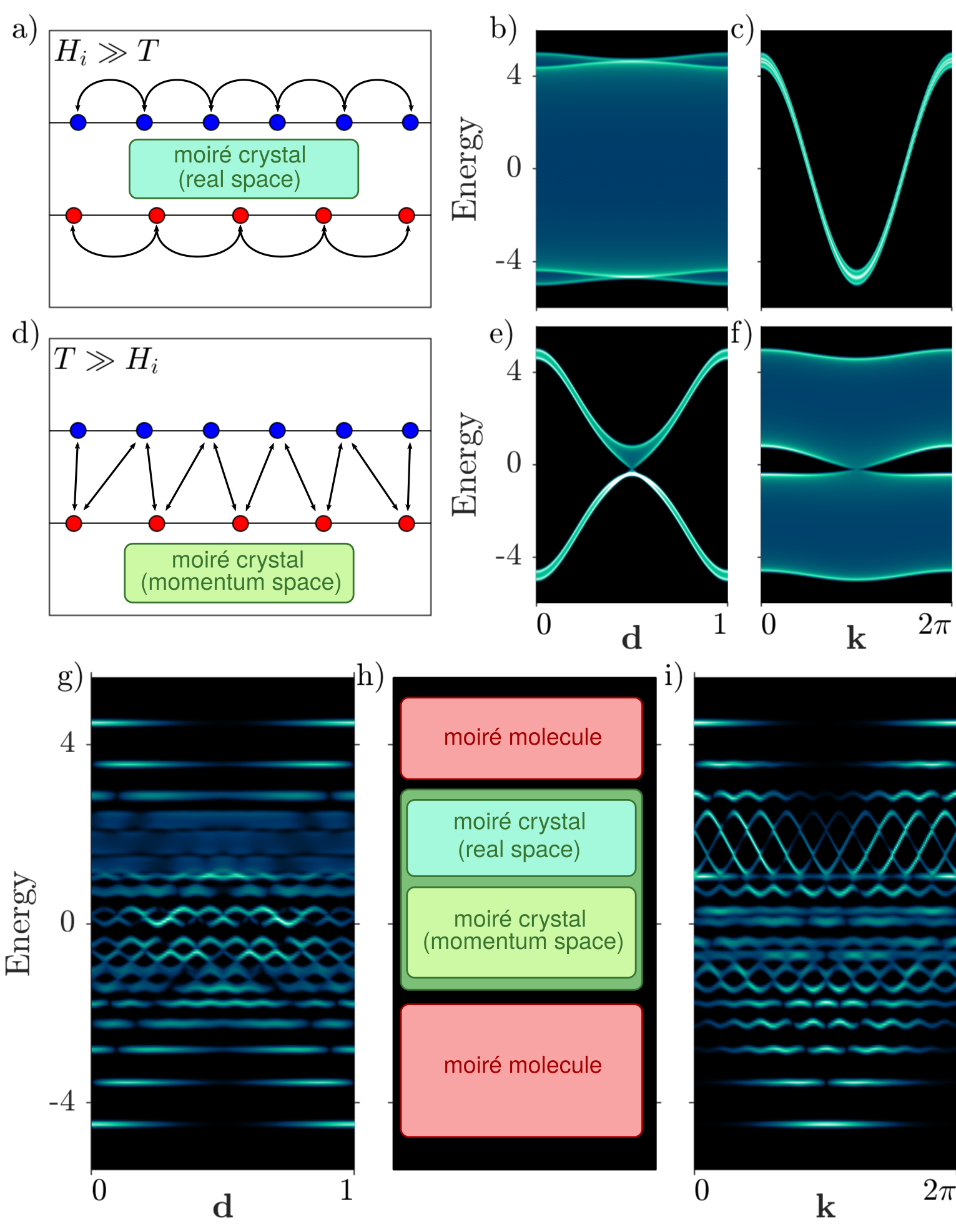}
      \caption{
            \textbf{a)} Diagram of the bilayer chain Hamiltonian when the monolayer couplings ($H_i$) are much larger than the interlayer coupling $(T)$.
            \textbf{b)} Configuration-dependent LDoS for the atoms of the top layer in this limiting case with $\Theta = 0$.
            \textbf{c)} Momentum-dependent LDoS for Bloch waves of the top layer in this limiting case with $\Theta = 0$.
            \textbf{c-f)} Same as \textbf{(a-c)}, but the interlayer coupling is now much larger than the monolayer couplings.
            \textbf{g-i)} Configuration and momentum LDoS for the bilayer chain model ($\Theta = 0.1$, $T_1 = 3$, $\xi = 0.3$) with the central panel describing the twistronic regime at that energy.
           }
      \label{fig:duality_crystal}
    \end{figure}

    The question arises, when and how does the previous expansion for $\omega_\Theta$ and the moir\'e molecule fail?
    Such a failure is visible in the low-energy regions of Fig. \ref{fig:duality_molecule} where the smooth density of a banded Bloch wave appears.
    This occurs exactly at the energy when the $\mathbf{d}$-dependent extremum in the LDoS reaches its minimum and changes direction.
    There is no longer any spatial confinement of the spectrum, but more importantly $T(\mathbf{d})$ can be treated as a constant in this energy range.
    From the Bloch-state point of view at this energy, $T(\mathbf{d})$ is constant (and thus can be neglected), reverting to the band structure of a typical 1D crystal.
    This is illustrated by the schematic picture and LDoS calculations in Fig. \ref{fig:duality_crystal}a-c.
    The standard result is obtained for the LDoS in configuration space (a continuum) and momentum space (a band).
    
    Connecting this to the derivation of $\omega_\Theta$, we see that the harmonic approximation fails because $\frac{\partial^2 T}{\partial \mathbf{d}^2}$ has vanished.
    A following question is, what if instead $\frac{\partial^2 E}{\partial \mathbf{k}^2}$ vanishes?
    This means the monolayer bands do not depend on $\mathbf{k}$ and the monolayer nearest-neighbor couplings are effectively zero.
    Such a scenario is shown in Fig. \ref{fig:duality_crystal}d-f.
    Comparing the two sets of LDoS calculations, we see the role of configuration and momentum have interchanged: the continuum of states is now in $\mathbf{k}$ while the band is in $\mathbf{d}$.
    This is still a crystal, but is unconventional in that its lattice occurs in the momentum basis and generates a band structure in configuration space.

    Taking $T_1 = 3 T_0$ and $\xi = 0.3$ (Fig. \ref{fig:duality_crystal}g-i) gives an interesting situation: both the conventional and unconventional crystals occur in the same spectrum.
    Isolated spots of density, associated with moir\'e flat bands, occur in both spaces.
    The two crystals occur, showing a continuous spectra in the original space and a clear band structure in its dual.
    Comparing this to the $\Theta = 0$ calculation of the same parameters in the lower-left panel of Fig. \ref{fig:1d_overview}c, the conventional crystal regime is predicted by a $\mathbf{d}$-independent energy region of the LDoS.
    For the unconventional crystal, there is an equivalent $\mathbf{k}$-independent energy region in the momentum-projected LDoS at $\Theta = 0$.

\section{Two-dimensional bilayers}
\label{sec:2d}
   
    We now turn our attention to moir\'e crystals in 2D.
    First, we will show the ubiquitous ability for the configuration-dependent LDoS in the untwisted geometry to predict localized electronic structures in twisted geometries by looking at examples of tBLG and tBLMoS$_2$.    Second, we will show how conventional band structure calculations of bilayer MoS$_2$ shed light on how the moir\'e molecule interpretation can describe the appearance and properties of moir\'e flat bands in generic twisted semiconductors. 
    
\subsection{Predicting twistronic features from LDoS}
    
    The three distinct regimes of electronic structure in the 1D model, the moir\'e molecule and the real and momentum space moir\'e crystal, can be directly predicted by careful interpretation of the untwisted LDoS.
    The moir\'e molecule, and its associated flat bands, occur when wells appear in the moir\'e potential.
    For a 2D crystal, this manifests in the LDoS as extrema or bright lines that have strong configuration dependence.
    The conventional moir\'e crystal, hosting traditional Bloch wave states, occurs in regions of the LDoS that have little configuration dependence.
    The unconventional moir\'e crystal and its configuration banding require a region of LDoS that has little $\mathbf{k}$ dependence after averaging over all configurations.
    This manifests in the configuration LDoS as regions with steep configuration dependence.
    
	\begin{figure*}
    	\centering
    	\includegraphics[width=\textwidth]{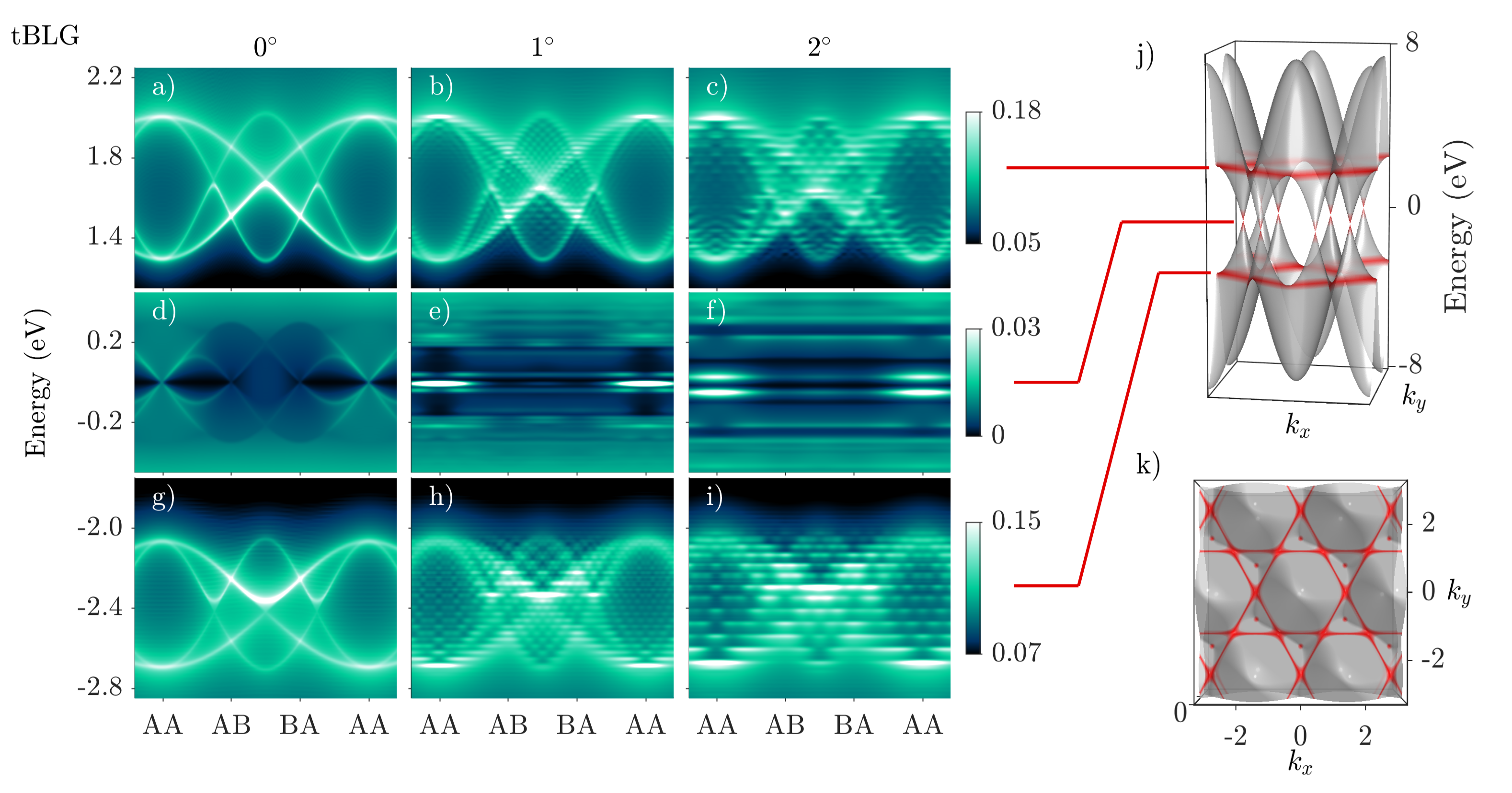}
    	\caption{
        \textbf{a-i)}
        Configuration-dependent LDoS for twisted bilayer graphene (tBLG) for selected energy regions and various twist angles.
    	The configurations correspond to a diagonal traversal of the moir\'e supercell.
    	The results were calculated for finite flakes of diameter $200$ nm with a Kernel Polynomial Method.
        \textbf{j)} Band structure for a simple nearest-neighbor tight-binding model of graphene.
        The energy regions associated with the selected LDoS energy regions of the twisted bilayer are highlighted in red.
        \textbf{k)} Top-down view of \textbf{(d)}.
        The Bloch states that are within the interlayer coupling strength (0.3 eV) of each selected energy are highlighted in red.
    	}
    	\label{fig:graphene_ldos}
    \end{figure*}
    
    Following these rules, the configuration dependence of the LDoS in untwisted bilayer graphene, Fig. \ref{fig:graphene_ldos}(a,d,g), can be used to predict all emergent electronic flat bands in the twisted cases.
    We first describe the important features of the untwisted electronic LDoS.
    Near the Fermi energy (Fig. \ref{fig:graphene_ldos}d), the LDoS shows conical features near $AA$ stacking.
    The parabolic dispersion of $AA$ bilayer graphene explains the bright features, and their configuration-dependence is due to band splitting as the sublattice symmetry is broken away from $AA$ stacking.
    Near  the $AB$ and $BA$ stackings, no bright features are visible, as the parabolic dispersion has disappeared.
    At energies far from the Dirac cones (Fig. \ref{fig:graphene_ldos}(a,g)), bright features in the LDoS also occur.
    They are caused by a band structure saddle point in the $p_z$-model for monolayer graphene, as illustrated in Fig. \ref{fig:graphene_ldos}j.
    As the Fermi energy moves through this saddle point, the curvature of the bands change sign and the Fermi surface undergoes a Lifshitz transition, causing a singular feature in the density of states.
    This singular feature (of the monolayer) is split by the effective interlayer coupling in the bilayer, and thus has strong configuration dependence.
    The largest splitting occurs at $AA$, with slightly weaker splitting at $AB$ and $BA$, consistent with the microscopic origins of the interlayer coupling.
    
    Looking at the $1^\circ$ twisted bilayer, a singular feature appears at $0$ eV and $AA$ stacking: this is  precisely the well-known magic angle flat band.
    It can be interpreted in the same fashion as solutions of the harmonic continuum model by replacing the quadratic monolayer Hamiltonians with the conical Dirac spinor \cite{Bistritzer2011}, which still provides analytic solutions \cite{Tarnopolsky2019}.
    At $2^\circ$, there are two bright features at low energy, and these correspond to the Van Hove singularities of the twisted bilayer.
    These features arise from the conical moir\'e potentials visible in the untwisted LDoS.
    
    For the two other energy regions, twisting causes isolated levels to occur in each of the $T(\mathbf{d})$ wells of the untwisted LDoS.
    In particular, between $AB$ and $BA$ a rather steep harmonic moir\'e potential well (relative to the monolayer band curvature) generates many stable harmonic oscillator levels, creating checkerboard patterns in the LDoS.
    Doubling the twist angle doubles the energy spacing between the states, showing that the harmonic approximation is still in good agreement with these features.
    Comparing the relative spacing between the states centered at $AA$ and the midpoint (between $AB$ and $BA$) we see that the curvature of $T(\mathbf{d})$ in the untwisted case proportionally affects the spacing between levels.
    This will be more thoroughly validated in the following subsection.

	\begin{figure*}
    	\centering
    	\includegraphics[width=1\linewidth]{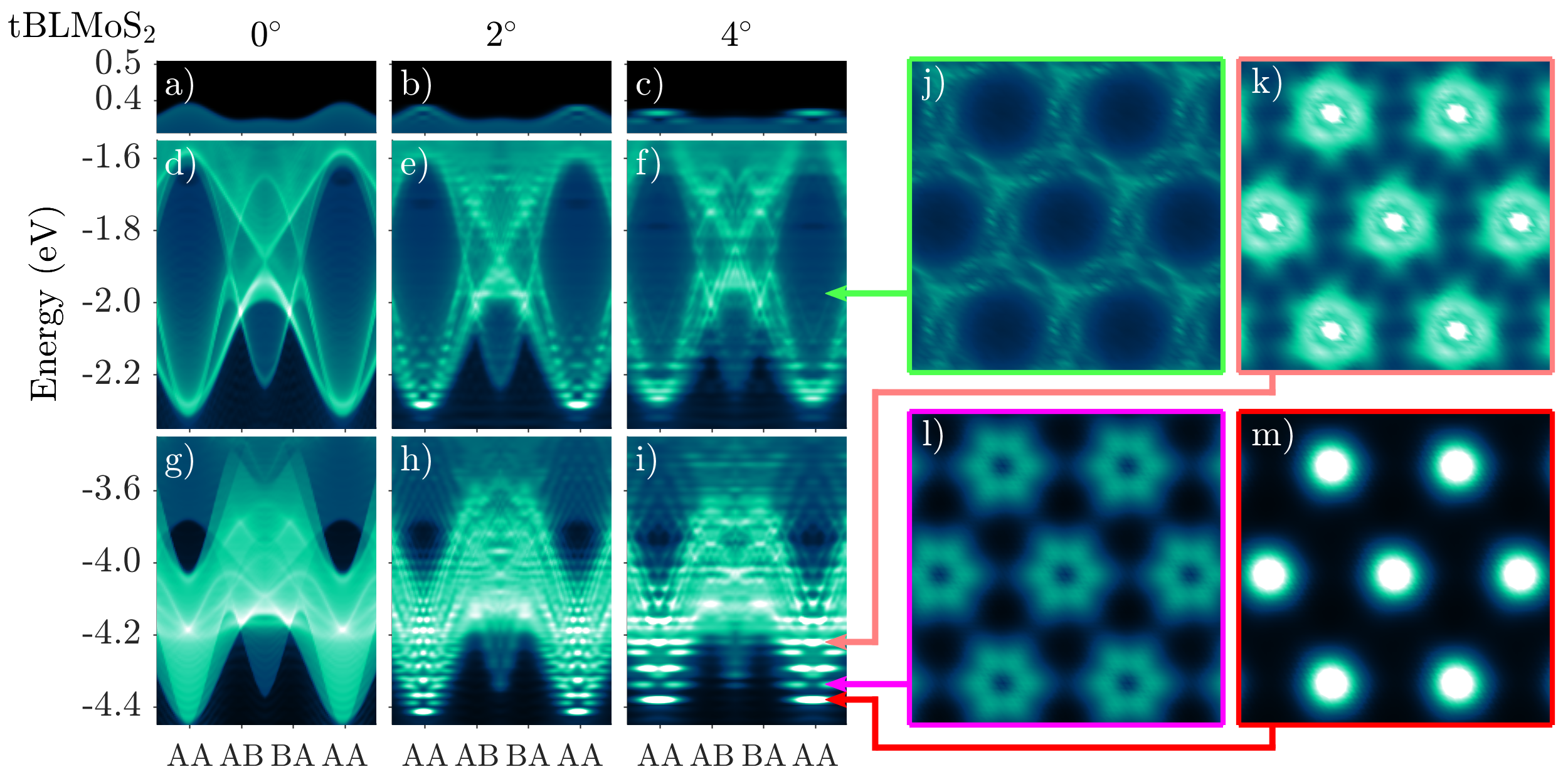}
    	\caption{
    	Configuration-dependent LDoS for the $p_z$ orbitals of bilayer MoS$_2$ in selected energy regions of the valence bands.
    	The configurations go across the diagonal of the moir\'e cell, starting and ending at aligned $AA$-type stacking regions.
    	Panels \textbf{(j-m)} show a top view of the configuration-dependent LDoS at four energy values for the $4^\circ$ twisted bilayer, interpolated from a $20 \times 20$ sampling of configuration space.
    	All results were calculated for finite flakes of radius 40 nm with the kernel polynomial method.
    	The energy spectrum resolution is limited by roughly 10 meV ($p = 4000$).
        }
    	\label{fig:tmdc_ldos}
    \end{figure*}
    
    Similar behavior can be observed in our calculations of twisted bilayer MoS$_2$ (tBLMoS$_2$).
    In Fig. \ref{fig:tmdc_ldos} we show the LDoS intensity of the $p_z$ orbitals of Sulfur as a function of energy and local atomic structure.
    These orbitals extend the farthest out of the plane of each layer and have the strongest interlayer coupling, so we have ignored other orbital contributions to the LDoS here.
    The other orbitals do couple weakly with the $p_z$ states in certain energy regions, and thus show faint signatures of twistronic features, but they primarily consist of a smooth LDoS background due to uninteresting Bloch waves (the conventional moir\'e crystal regime).
    
    The first energy region is the valence band maximum (Fig. \ref{fig:tmdc_ldos}a-c).
    The LDoS shows a step-function behavior in the vertical (energy) direction, consistent with the density of states of two-dimensional parabolic bands.
    At $4^\circ$, signatures of moir\'e flat bands are visible, but hard to resolve as the KPM broadens the eigenvalues by a Gaussian of width $\sigma = 5$ meV in these calculations.    The next energy region (Fig. \ref{fig:tmdc_ldos}d-f) has singular features in the LDoS.
    This region arises from a Lifshitz transition in a $p_z$-majority band, and is functionally identical in shape to the feature in the graphene model.
    This implies that this particular twistronic structure is a signature of $p_z$ bands on a triangular lattice.
    As it occurs in both graphene and MoS$_2$, the structure does not seem to require sublattice symmetry on the honeycomb lattice.
    Similar to graphene, moir\'e molecular states are visible at each of the band extrema.
    In contrast, weak configuration banding is visible near the center of the feature ($-2.0$ eV), which was missing in tBLG, due to stronger interlayer coupling relative to the monolayer band curvature in the TMDCs.
    This agrees with the requirement of a sufficiently strong interlayer moir\'e potential in the 1D model (Fig. \ref{fig:1d_overview}c).
    
    The last energy region (Fig. \ref{fig:tmdc_ldos}g-i) again shows LDoS features consistent with a 2D band extremum, however the configuration dependence is much stronger than at the valence band maximum.
    Accordingly, clear moir\'e molecular states exist near $-4.4$ eV, and more robust configuration banding occurs at the center ($-4.0$ eV).
    Two independent sets of $AA$-centered harmonic oscillator states appear more clearly in this last case, owing to the two $AA$-centered $T(\mathbf{d})$ wells at $-4.4$ eV with different curvature.
    The LDoS corresponding to the $n = 0$ and $n = 1$ harmonic energy levels of the shallower moir\'e potential are shown in Fig. \ref{fig:tmdc_ldos}l,m displaying $s$ and $p_x \pm i p_y$ symmetries, respectively.
    In Fig. \ref{fig:tmdc_ldos}k a bright center inside a ring is visible, a signature of $d$ orbital momentum from the $n = 2$ level of the sharper well.
    At slightly higher energy ($-4.3$ eV) a $T(\mathbf{d})$ well centered in between $AB$ and $BA$ also hosts moir\'e molecular states, although they are not as well confined.
    The coexistence of multiple unique moir\'e flat band modes at the same energy could give rise to unique quantum phases.
    
    These results introduce an additional constraint necessary in the classification of possible twistronic phenomena of 2D materials: the connectivity of the constituent monolayer band structures.
    The moir\'e molecule and its associated flat bands requires a small number of connected Bloch states of similar energy, and therefore it naturally occurs in parabolic band extrema and Dirac points.
    The momentum space crystal can occur at 2D Lifshitz transitions in the monolayer band structure.
    A previous work \cite{Massatt2018} showed that these transitions can create an infinite lattice of coupled Bloch states, preventing convergence of the truncated continuum models for twistronic systems.
    This describes the formation of a momentum space crystal, albeit only on a subset of the Brillouin zone, and the failure of the continuum models is directly related to the breakdown of the dual nature of the moir\'e molecule.
    This means saddle points in the 2D bandstructure are good candidates for the realization of configuration banding.
    An alternate approach would be to find a system where the interlayer coupling $T$ is larger than the entire width of a monolayer band.
    In van der Waals materials, the interlayer distance between orbitals is larger than the in-plane distances, so this is a challenging task.
    
    The Lifshitz regions also host robust moir\'e flat-bands at the interlayer coupling extrema, which justifies experimental study of twisted bilayers made of metallic materials.
    Unlike flat bands arising from Dirac cones or band extrema, it seems necessary for these to exist within a background of metallic states.
    The LDoS calculations are particularly useful for identifying these modes, as band structure calculations are difficult to interpret within a metallic background and continuum methods are unlikely to converge \cite{Massatt2018}.

    \begin{figure*}
      \centering
      \includegraphics[width=1\linewidth]{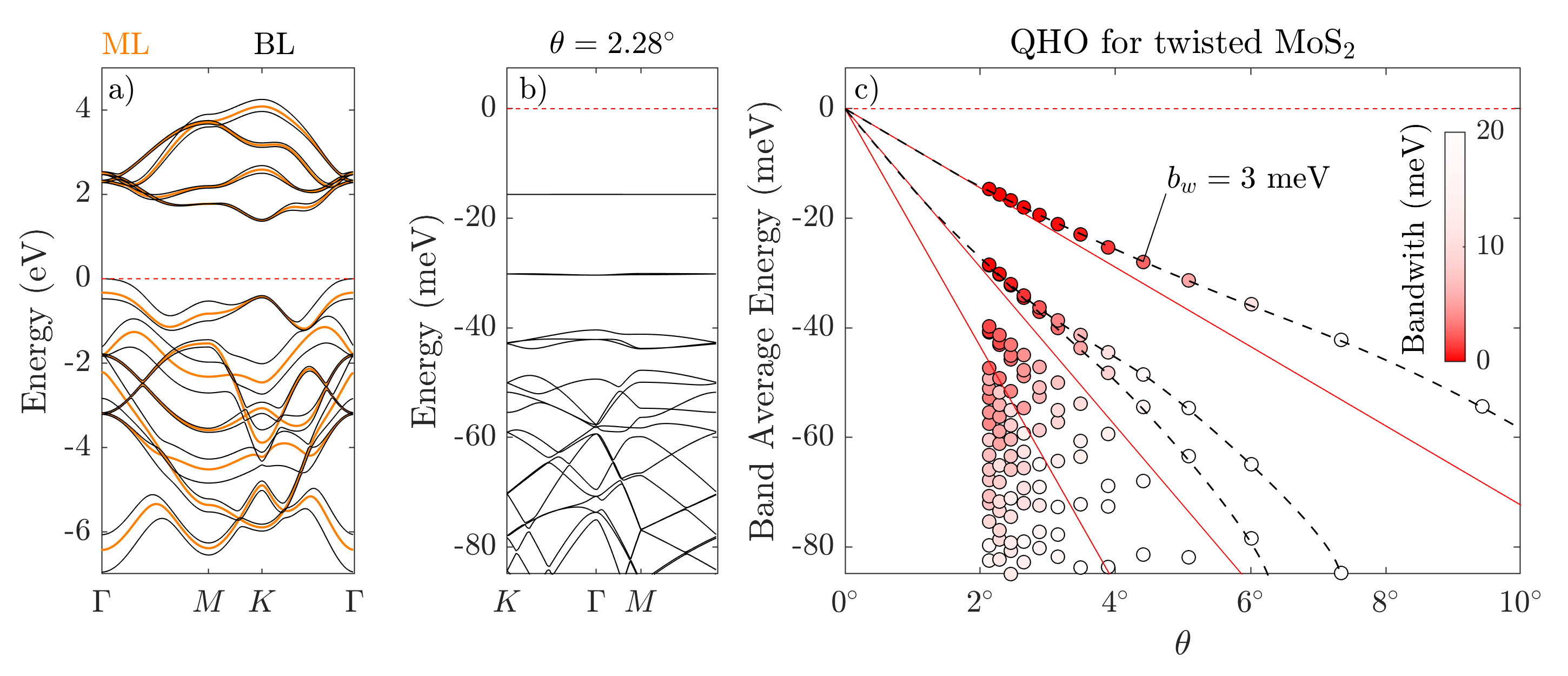}
      \caption{Interpretation of the moir\'e flat bands in unrelaxed twisted bilayers of MoS$_2$ as a quantum harmonic oscillator model.
      \textbf{a)} Band structure comparison between a monolayer (orange) and $AA$ stacked bilayer (black) of MoS$_2$.
      The effective interlayer coupling can be easily interpreted at any band and momentum by looking at the strength of the band splitting.
      The dashed red line indicates the bilayer's valence band maximum, which is the 0 energy reference for the following panels.
      \textbf{b)} Band structure of the twisted bilayer at $\theta = 2.28^\circ$, showing the characteristic splitting and flattening of the bands, as captured by the QHO model.
      \textbf{c)} Summary of band structures for twisted bilayer MoS$_2$ near the valence band maximum.
      The average band energy for each band is plotted as a function of the twist angle $\theta$, with the bandwidth $b_w$ given in color.
      The red lines indicate the expected energy levels of a harmonic oscillator defined by the monolayer's effective mass $m^*$, and the shape of the moir\'e potential $T(\vec{r})$,  both expanded around $\Gamma$.
      The dashed lines are a guide to the eye for $\theta$ dependence of the first (singly degenerate) level and the second (double degenerate) level.}
      \label{fig:QHO}
    \end{figure*}
    
\subsection{Flat bands in MoS$_2$}

    The harmonic approximation of band extrema for the incommensurate 1D chain is a general interpretation of any moir\'e flat bands arising from interlayer hybridization.
    This approximation can also give reliable predictions of band flattening at the valence band maximum of MoS$_2$.
    These results are in agreement with the LDoS calculations of the previous section, but by performing band structure calculations the energy resolution limitation is removed, allowing for careful comparison to the harmonic oscillator.
    This band extremum (maximum) 
    is at $\Gamma$ and is maximized  at $AA$ stacking, as shown in Fig. \ref{fig:QHO}a.
    Accordingly, we take this maximum value as $E = 0$.
    Following the 1D case, we then extract the effective mass $m^*$ from a monolayer calculation and the configuration dependence of the $\Gamma$ band extremum energy, which provides the effective moir\'e potential $T(\mathbf{d})$.
    Both are obtained by fitting up to the quadratic term to extract the effective harmonic frequency parameter, $\omega_\theta = -7.24$ meV per degree.
   	For a given a twist angle, the harmonic frequency is given by $\omega = \omega_\theta \theta$.

    The band structure calculations show excellent agreement to the energy levels of the effective harmonic oscillator, $\omega (n + 1)$.
    Unlike the 1D case, the $n^\textrm{th}$ 2D harmonic oscillator level is $(n+1)$-fold degenerate.
    Looking at the band structure of the $\theta = 2.28^\circ$ supercell ($\lambda = 8$ nm) in Fig. \ref{fig:QHO}b, we find a single flat band with nearly zero bandwidth at the expected value of $\omega_\theta \theta = -16.5$ meV. 
    Lower in energy, a pair of nearly degenerate flat bands have a value just shy of the next energy level, $2 \omega_\theta \theta = -33$ meV.
    Below this, a set of three bands have separated from the remainder of the valence band manifold, although they still have a substantial bandwidth.
    This analysis is performed for each supercell and we report the band average energy and total bandwidth along the high-symmetry lines in Fig. \ref{fig:QHO}c.
    Near $2^\circ$, the first flat band is in perfect agreement with the QHO model, but as $\theta$ increases the band becomes less flat and its average begins to drift from the QHO model.
    This same behavior can be seen for the $n=1$ and $n=2$ levels, and so in the $\theta \to 0^\circ$ limit we expect every $n$ will emerge as a flat band in this simple tight-binding model.
    Following a similar constraint of the band flattening in the 1D model, we find the $n^\textrm{th}$ level will become sufficiently flat ($b_w < 5$ meV) when the $(n+2)^\textrm{th}$ level is contained within the moir\'e potential well.
    The moir\'e potential for the MoS$_2$ model used here is approximately 100 meV deep. 
    That means the critical twist angle for the $n = 0$ flat band should satisfy $| (2+1) \omega_\theta \theta^{(0)}_c | < 100$ meV, yielding $\theta^{(0)}_c \approx 4.6^\circ $.
    For the second and third level of the oscillator, this similarly provides critical values of $\theta^{(1)}_c \approx 3.5^\circ$ and $\theta^{(2)}_c \approx 2.8^\circ$.
    These are all in excellent agreement with the results of the tight-binding calculations, namely that when $\theta < \theta^{(n)}_c$ the bandwidth for the $n^\textrm{th}$ level is below $5$ meV.

    Comparing this result to full DFT calculations \cite{Naik2018} we find good agreement to the unrelaxed case: the flat-bands are associated with $AA$ stacking spots at the valence band maximum.
    However, the inclusion of atomic relaxations changes the valence flat band characteristics.
    This is primarily due to large variations in the interlayer separation as a function of atomic configuration \cite{Naik2018,Xian2019}, which significantly modifies the interlayer moir\'e potential.
    The predicted moir\'e potential also depends sensitively on the choice of van der Waals DFT functional \cite{Zhou2015}.
    For this reason, the results presented in Fig. \ref{fig:QHO} are meant to serve as an introduction to the band flattening phenomena in twisted semiconductors, not as a fully self-consistent prediction of flat bands.
    See Refs. \onlinecite{Naik2018, Xian2019} for fully \textit{ab-initio} predictions of band flattening in semiconductors, including important relaxation effects.
    
    Although the derivation of an explicit harmonic oscillator equation relies on the maximum being at $\Gamma$, it is not a necessary condition for the harmonic oscillator interpretation of the flat bands.
    Any band extremum in any material can host moir\'e flat bands, with its own characteristic twist angle depending on the interlayer coupling strength and effective mass.
    From this perspective, the Dirac cones of graphene are an exception to the rule.
    The two-sided nature of the Dirac equation causes the flat-band condition to occur only under a fine tuning of the twist angle to a magic value.
    For generic insulators or semiconductors this fine tuning is not necessary, as reducing the twist angle always causes a shallower moir\'e potential and leads to more confined harmonic states and thus flatter bands \cite{Wang2019}.
    
    In experiment one wants not only to minimize the bandwidth, $b_w$, but also to keep each set of flat bands separated sufficiently in energy.
    In the unrelaxed model studied here, it is clear that although one can make the bandwidth arbitrarily small as $\theta \to 0^\circ$, the band-gaps (given by $\omega_\theta \theta$) will vanish.
    A tradeoff between band width and band gap should be considered depending on the goal of the proposed experiment.
    For angles well above the first critical angle, $\theta^{(1)}_c \approx 4.6^\circ$, no bands are guaranteed to be flat or gapped from the rest of the valence manifold.
    If more than one harmonic oscillator level is required, or a certain minimum bandwidth is desired, this will set an upper bound for the desired twist angle.
    In both cases any constraint on the minimum allowed band gap will set a lower bound on the twist angle.
    Unlike twisted bilayer graphene, variations in the twist angle \cite{Uri2020} are not likely to fundamentally alter the correlated behavior of electrons in the material, as the band flatness is generally monotonic with respect to the twist angle.
    
    Additional corrections should be carefully considered in the future.
    The anharmonicity in the effective mass and interlayer coupling will be necessary for accurate comparison to experimental results, but the overall intuition for the moir\'e flat bands is unchanged.
    In addition, in-plane relaxations at small twist angles form domain-wall structures \cite{Carr2018relax, Zhang2018, Yoo2019, Rosenberger2020} and will cause the moir\'e potential to become independent of the twist angle \cite{Carr2019cont}.
    This prevents the tuning of $\omega$ to arbitrarily small values, and will define a minimum possible band width for the flat bands.
    As the $\Gamma$ bands of MoS$_2$ are primarily $p_z$ character and thus have negligible spin splitting, spin-orbit coupling was not needed here.
    If the moir\'e bands instead arise from the $K$ valley in a TMDC material other than MoS$_2$, spin-orbit coupling can be important \cite{Wu2017, Wu2018, Wang2019}.

\section{Conclusion}
\label{sec:conc}

    We have provided a classification of twist induced electronic structure in 1D and 2D moir\'e systems through the local density of states.
    The self-dual nature of the moir\'e Hamiltonians puts the momentum and configuration bases on equal footing, providing insight into twistronic features.
    In particular, moir\'e flat bands can be robustly predicted from either the LDoS or band structures of untwisted systems.
    The LDoS approach also allows for prediction of flat bands even in metallic backgrounds, and explains the formation of effective momentum crystals, problems not yet addressed by band structure methods.
    The study of moir\'e heterostructures made of arbitrary 2D materials is made much easier within this framework.
    Apart from the strengthening of correlated effects, many applications that take advantage of the electronic localization in these structures can be envisaged, such as arrays of quantum dots or networks of 1D electron channels.

\begin{acknowledgements}
We thank Shiang Fang, Paul Cazeaux, and Zoe Zhu for helpful discussions. 
This work was supported by the STC Center for Integrated Quantum Materials NSF Grant No. DMR-1231319, NSF DMREF Award 1922165, and ARO MURI Award W911NF-14-0247. 
\end{acknowledgements}

\appendix*
\section{Configuration bands and the moir\'e comb}

	\begin{figure*}
    	\includegraphics[width=1\textwidth]{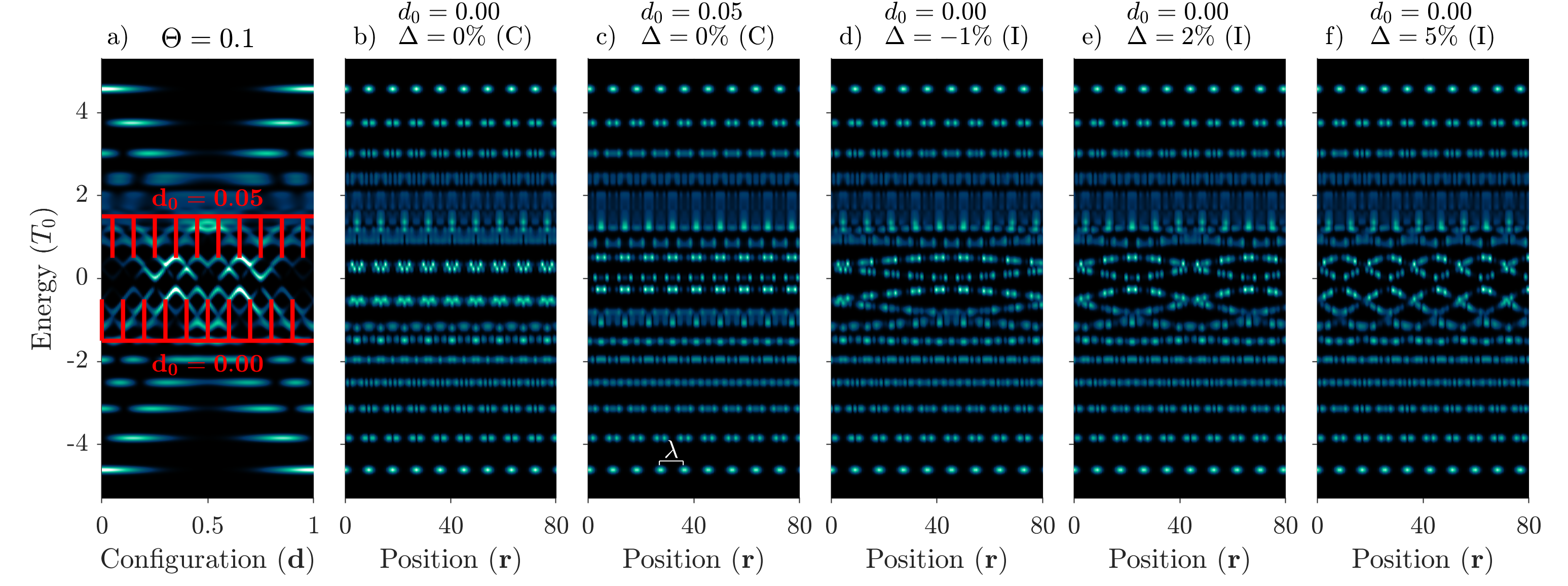}
    	\caption{ 
    	\textbf{a)} LDoS for the 1D bilayer model as a function of local configuration $\mathbf{d}$ at $\Theta = 0.1$.
    	For a commensurate geometry only a finite number of configurations are sampled.
    	The sampled configurations for two different choices of $\mathbf{d}_0$  are indicated by the red ``moir\'e combs''.
    	\textbf{b-c)} Real space LDoS for the two commensurate geometries (C) introduced in \textbf{(a)}.
    	The configuration banding is lost due to the finite sampling of configurations.
    	The moir\'e length $\lambda$ is given at the bottom of $\mathbf{(c)}$.
    	\textbf{d-f)} Real space LDoS for three selected incommensurate geometries (I), parameterized by $\Theta = 0.1 \times (1 + \Delta)$.
    	As these geometries are not periodic on the moir\'e length, the configuration bands are visible in realspace.
    	}
    	\label{fig:comb}
    \end{figure*}

    The configuration bands of the momentum space moir\'e crystal are clear in the configuration dependent LDoS of the 1D model (Fig. \ref{fig:duality_crystal}g), but one may ask what these electronic states look like in real space.
    Calculations of the LDoS for a system with a momentum space crystal are presented in Fig. \ref{fig:comb} for various choices of $\Theta$.
    It begins with a commensurate cell, $\Theta_0 = 0.1$.
    
    To understand the role of incommensurate systems in the context of configuration banding we introduce incommensuration through the parameter $\Delta$, setting $\Theta = \Theta_0 \times (1 + \Delta)$.
    $\Delta$ must be irrational to generate a truly incommensurate system.
    However, in calculating the $\mathbf{d}$-dependent LDoS via the finite-sized method (for both 1D and 2D bilayers), we find that small values of $\Delta$ have no qualitative effect when $\Theta_0$ is small.
    The $\mathbf{d}$-dependent LDoS is a stable quantity under small variations in the twist angle.
    For this reason, in the following discussion and Fig. \ref{fig:comb} we always use the configuration LDoS calculation for $\Theta = \Theta_0$, and have $\Delta$ only affect the sampling of $\mathbf{d}$ to generate the real space LDoS.
    
    The interlayer configuration at the origin ($\mathbf{r} = 0$) is controlled by the parameter $\mathbf{d}_0$.
    For commensurate systems, this choice of $\mathbf{d}_0$ is only unique up to $1/\lambda$, as units of the inverse moir\'e length correspond to a translation of the origin.
    Modifying $\mathbf{d}_0$ will show the relative stability of the real space LDoS under small changes in interlayer alignment in commensurate systems.

    For all choices of $\Delta$, the high and low energy regions show isolated electronic states that are robust under changes in the initial configuration $\mathbf{d}_0$ and incommensuration $\Delta$. 
    These are the harmonic oscillator states of the moir\'e molecule, and their stability with respect to $\Delta$ is explained by the insensitivity of the previous harmonic expansion to the microscopic details of atomic geometry.
    
    For commensurate geometries, a slight relative movement of the layers can drastically alter the electronic structure in the momentum space crystal regime.
    Near 0 energy in Fig. \ref{fig:comb}b,c the energies of the localized modes change significantly when the configuration is moved by half of $\Theta$.
    This is because a commensurate system only samples a finite number of configurations, and so changing the initial configuration $\mathbf{d}_0$ causes different sections of the configuration bands to manifest.
    A visual aid for inferring the real space LDoS from the configuration space LDoS is a ``moir\'e comb'' which has a spacing $\Theta$ between each tooth, as shown in Fig. \ref{fig:comb}a.
    The two red combs correspond to two values of $\mathbf{d}_0$ which generate geometries with vastly different LDoS near 0 energy.
    Changing the initial configuration of the commensurate system corresponds to moving the comb left or right, while changing the energy of interest moves it up or down.
    The electronic structure of these commensurate geometries in the momentum crystal energy range looks markedly similar to the stable isolated states of the moir\'e molecule, but they are highly sensitive to small changes in alignment between the layers.
    Care must be taken in moir\'e supercell calculations to avoid interpreting these features as stable flat bands.

    Incommensurate geometries, or geometries which are not periodic on the moir\'e length, with $\Delta \neq 0$ can be interpreted as a sequence of moir\'e supercells of length $1/\Theta_0$ but with the variable $\mathbf{d}_0$ varying in space.
    The isolated electronic states that result band with a wavelength determined by a second-order moir\'e length, which is the length required for $\mathbf{d}_0$ to go once through the range $[0, L]$.
    This electronic structure hosts highly localized wavefunctions, which are not directly provable from the LDoS calculations, but have been studied extensively for incommensurate potentials acting on 1D tight-binding models \cite{Aubry1980,Grempel1982,Kohmoto1983}.
    In particular, self-dual models of this form \cite{Aubry1980,Johansson1991} bear strong resemblance to the momentum-space crystal structure studied here.
    Making a direct connection between the moir\'e system and these incommensurate potential models via explicit comparison of the tight-binding models is difficult, owing to the inability to pair atoms of the top and bottom layer together in a consistent manner across the surface of a moir\'e pattern \cite{Balents2019}.
    However the previous $\psi_\pm$ model (Eq. \ref{eq:psi_pm_model}) is identical to the incommensurate potential models at the continuum limit, provided one replaces the $\mathbf{k}^2$ term with a complete monolayer Hamiltonian.

\bibliography{config_momentum_duality}

\end{document}